\documentclass[aps,prd,a4paper,onecolumn,amsmath,showpacs,superscriptaddress,nofootinbib,preprintnumbers,notitlepage]{revtex4-1}

\usepackage{verbatim}
\usepackage[T1]{fontenc}
\usepackage[utf8]{inputenc}
\usepackage[american]{babel}
\usepackage{epsfig}
\usepackage{graphicx,subcaption,caption}
\usepackage{pgfplots}
\usepackage{booktabs}
\usepackage{multirow}
\usepackage{dcolumn}
\usepackage{amsmath}
\usepackage{mathtools}
\usepackage{amsfonts}
\usepackage{amssymb}
\usepackage{epstopdf}
\usepackage{bm}
\usepackage{siunitx}
\usepackage{braket}
\usepackage{enumitem}
\usepackage{soul}
\usepackage{ulem}
\usepackage{color}
\usepackage{transparent}
\usepackage{pifont}
\usepackage[font={small}]{caption}

\definecolor{navyblue}{rgb}{0.0, 0.0, 0.5}
\definecolor{royalblue}{rgb}{0.25, 0.41, 0.88}
\definecolor{cadmiumgreen}{rgb}{0.0, 0.42, 0.24}
\definecolor{blue-violet}{rgb}{0.54, 0.17, 0.89}
\definecolor{darkviolet}{rgb}{0.58, 0.0, 0.83}
\definecolor{orange(colorwheel)}{rgb}{1.0, 0.5, 0.0}

\usepackage{hyperref}
\hypersetup{
    colorlinks=true, 
    linkcolor=royalblue, 
    citecolor=magenta}

\newcommand\be{\begin{equation}}
\newcommand\ee{\end{equation}}

\newcommand\bea{\begin{eqnarray}}
\newcommand\eea{\end{eqnarray}}

\renewcommand{\vec}{\bm}

\usepackage{booktabs}
\usepackage{multirow}
\usepackage{dcolumn}
\usepackage{colortbl}

\definecolor{magenta(process)}{rgb}{1.0, 0.0, 0.56}

\definecolor{darkspringgreen}{rgb}{0.09, 0.45, 0.27}

\definecolor{royalblue(web)}{rgb}{0.25, 0.41, 0.88}

\begin{document}

\title{Weak Gravity Conjecture from Conformal Field Theory: A Challenge from Hyperscaling Violating and Kerr-Newman-AdS Black Holes}

\author{Jafar Sadeghi}
\email{pouriya@ipm.ir}
\affiliation{Department of Physics, University of Mazandaran, P. O. Box 47416-95447, Babolsar, Iran}

\author{Mehdi Shokri}
\email{mehdishokriphysics@gmail.com}
\affiliation{School of Physics, Damghan University, P. O. Box 3671641167, Damghan, Iran}
\affiliation{Department of Physics, University of Tehran, North Karegar Ave., Tehran 14395-547, Iran}
\affiliation{Canadian Quantum Research Center 204-3002 32 Avenue Vernon, British Columbia V1T 2L7 Canada}

\author{Mohammad Reza Alipour}
\email{mr.alipour@stu.umz.ac.ir}
\affiliation{Department of Physics, University of Mazandaran, P. O. Box 47416-95447, Babolsar, Iran}

\author{Saeed Noori Gashti}
\email{saeed.noorigashti@stu.umz.ac.ir}
\affiliation{Department of Physics, University of Mazandaran, P. O. Box 47416-95447, Babolsar, Iran}

\preprint{}
\begin{abstract}
We search for a possible relation between weak gravity conjecture and conformal field theory in the hyperscaling violating and Kerr-Newman-AdS black holes. We deal with the critical points of the black hole systems through the correlation function introduced in conformal field theory. We discuss the weak gravity conjecture conditions by the imaginary part of the energy obtained from the critical points and their poles. Under the assumptions $z=1$, $d=1$ and $\theta\rightarrow0^{-}$,  we link weak gravity conjecture to the hyperscaling violating black holes due to the existence of $r_{H}$ larger and smaller than one. As the second black hole system, we study the conditions of weak gravity conjecture for the Kerr-Newman-AdS black hole by rotation and radius parameters. Then, we show that the conditions of weak gravity conjecture are satisfied when the charged particle near the hyperscaling violating and Kerr-Newman black holes is $\frac{1}{a}$ with a ratio $\frac{a}{\ell}\ll 1$.
\end{abstract}

\maketitle

\section{Introduction}\label{into}

The AdS/CFT correspondence has been known as one of the essential dualities between quantum field theory (QFT) and gravity that corresponds to classical dynamics of gravity in a higher dimension. The duality was first introduced to connect four dimensions conformal field theory to the Anti-de Sitter space (AdS) in five dimensions \cite{Maldacena:1997re,Gubser:1998bc,Witten:1998qj}. The AdS/CFT correspondence has been applied to a wide range of physical science e.g. strong-coupling dynamics (QCD), the physics of black holes and quantum gravity, electroweak theories, relativistic hydrodynamics \cite{Klebanov:1999tb,Kioumarsipour:2021zyg,Bu:2021jlp,Fujiwara:2021xgu,Evans:2021zzm,MartinContreras:2021bis,Gherghetta:2009ac,Brodsky:2007hb,Nakano:2006js,Katz:2005ir,Meltzer:2019nbs,Karch:2006pv,Andreev:2006ct,Cavaglia:2021mft,Harmark:2020vll,BitaghsirFadafan:2020lkh,DeLeeuw:2020ahx,DeWolfe:2020uzb,Ishigaki:2020vtr,Terashima:2020uqu,Yin:2021zhs,Mes:2020vgy,Berenstein:2020cll,Berman:2022idl,Aharony:1999ti,DHoker:2002nbb,Hartnoll:2009sz,McGreevy:2009xe,Kim:2012ey,Adams:2012th}.

One of the most significant challenges in theoretical particle physics is finding an effective field theory (EFT) compatible with quantum gravity. Such theories are situated in the landscape while other incompatible EFTs live in the swampland. A universal test to distinguish between these two classes of theories is weak gravity conjecture (WGC).  This means that gravity is always the weakest force and shows an extremality state of the black hole  \cite{Vafa:2005ui,Ooguri:2006in,Arkani-Hamed:2006emk,Sadeghi:2021plz,Cordova:2022rer,Henriksson:2022oeu,Kaya:2022edp,Collazuol:2022jiy,Capozziello:2011nr,McInnes:2022tut,Cheung:2014vva,Rudelius:2022gyu,Cribiori:2022trc,Klaewer:2020lfg,Heidenreich:2015nta,Heidenreich:2016aqi,Andriolo:2018lvp,Ooguri:2006in,Polchinski:2003bq,Ooguri:2016pdq,Ibanez:2017kvh,Shiu:2016weq,Fisher:2017dbc,Cheung:2018cwt,Crisford:2017gsb,Harlow:2015lma,Hamada:2018dde,Kinney:2018nny,Harlow:2022gzl,Saraswat:2016eaz,Nakayama:2015hga,Bachlechner:2015qja,Bellazzini:2019xts,Aharony:2021mpc,Heidenreich:2019zkl,Benakli:2020vng,Lee:2018spm,deAlwis:2019aud,Andriolo:2020lul,Cremonini:2020smy,Ibanez:2015fcv,Banerjee:2020xcn}.

In the present manuscript, we attempt to prove WGC from CFT by studying two critical classes of black holes e.g. hyperscaling violating (HSV) and Kerr-Newman-AdS (KNA) black holes. Since WGC is more consistent at critical points, we use CFT calculations in order to obtain the critical points associated with the black holes under consideration. The above discussion motivates us to arrange the paper as follows. In section \ref{wgc}, we briefly explain how the WGC could be related to the CFT. In section \ref{hsv}, we investigate the relationship between the WGC and the CFT in the HSV black holes. We perform the similar analysis for the KNA black holes in section \ref{knb}. Conclusions and remarks are drawn in section \ref{con}.

\section{A bridge between WGC and CFT}\label{wgc}
The swampland suffers from the lack of global symmetries and completeness of the charge spectrum. Consequently, it cannot present an acceptable explanation for phenomena unless we restrict the global symmetry and whether there exists an upper bound on the mass of some charged conditions \cite{Heidenreich:2021yda,Fichet:2019ugl,Daus:2020vtf,Kats:2006xp,vanBeest:2021lhn}. The swampland can restrain a complete theory, not EFTs with low energy. Phenomenologically, it is important to answer whether all charged particles are heavy and actual enough concerning the black holes or if there exist several concepts of the spectrum completeness that prevails at low energies. The swampland conjectures discuss some important issues, such as they aspire to point out how near we can obtain to the status of recovering the global symmetries. One can refresh a global symmetry of $U(1)$  by transmitting the gauge coupling to 0, which should not be permitted in quantum gravity.

Attempting to comprehend how string theory prohibits this issue and what drives incorrect if one tries to accomplish this can supply facts about the restrictions that EFT must meet to be compatible with quantum gravity. WGC prohibits this methodology by the existence of unique light-charged forms that exclude the definition of EFT. This supplies an upper bound on the mass of charged states. In general, WGC includes the magnetic and electric versions providing a gauge theory coupled to gravity. So we deal with electrically charged Planck units as $Q/m\geq\mathcal{Q}/M |_{ext}=\mathcal{O}(1)$ \cite{Heidenreich:2021yda,Fichet:2019ugl,Daus:2020vtf,Kats:2006xp,vanBeest:2021lhn}. Here, $Q$ is charge, and $M$ determines mass in the extremal black hole. Also, $Q=qg$, where $q$ and $g$ are the quantized charges of the state and the gauge coupling, respectively.

To have WGC with the existence of charge and mass need to apply the charge/mass greater than the one in the extremal black hole. The most specific topic coordinates with the theory of Maxwell, which is coupled to gravity; no massless scalar fields exist. So, we can construct solutions of R-N black hole, so provided a $p$-form gauge field in d dimensions. WGC indicates the presence of a ($p-1$)-brane can meet the $ p(d-p-2)T^{2}/d-2 \leq Q^{2}M_{p}^{d-2}$ \cite{Harlow:2022gzl,Saraswat:2016eaz,Nakayama:2015hga,Bachlechner:2015qja,Bellazzini:2019xts,Aharony:2021mpc,Heidenreich:2019zkl,Benakli:2020vng,Lee:2018spm,Daus:2020vtf,Katz:2005ir,vanBeest:2021lhn}. Here we note that the motivation for understanding WGC is well seen in the physics of black holes. This conjecture expresses that the entire lattice of authorized gauge charges must be settled by physical conditions in an approach with a gauge coupled to gravity. This is not required in quantum field theory since a charged particle can be decoupled from this theory by transmitting the mass to $\infty$. The second issue is breaking global symmetries. There are fascinating relationships between the lack of global symmetries in quantum gravity and spectrum completeness. A typical method of breaking higher form global symmetries includes charged conditions. While exclusively if the charged conditions spectrum exists entire, can one break the whole group\cite{Arkani-Hamed:2006emk,Sadeghi:2021plz,Cordova:2022rer,Henriksson:2022oeu,Kaya:2022edp,Collazuol:2022jiy,Capozziello:2011nr,McInnes:2022tut,Cheung:2014vva,Harlow:2022gzl,Saraswat:2016eaz,Nakayama:2015hga,Bachlechner:2015qja,Bellazzini:2019xts,Aharony:2021mpc,Heidenreich:2019zkl,Benakli:2020vng,Lee:2018spm}.

In this article, we want to prove or emerge WGC from CFT equations. Hence, we discuss the mixed Klein-Gordon equation against the background of the black hole as a general perturbation, and we focus on a charged scalar $\Phi$  with charge $q$ and mass $m$ \cite{Konoplya:2013rxa}
\begin{equation}\label{eq1}
\frac{1}{\sqrt{-g}}\partial_{\mu}(g^{\mu\nu}\sqrt{-g}\partial_{\nu}\Phi)-2iq g^{\mu\nu}A_{\mu}\partial_{\nu}\Phi-q^2 g^{\mu\nu}A_{\mu}A_{\nu}\Phi-m^2\Phi=0.
\end{equation}
By inserting $\Phi=e^{-i\omega t} \phi(\vec{x})$, we get its scalar function also $\omega$ is related to the energy part. Then, we use CFT to find the two-point correlation function of the scalar operator $J_k$. We also can find the correlation function based on the ratio of its sub-coefficients as follows. \cite{Urbano:2018kax,Alishahiha:2012ad}
\begin{equation}\label{eq2}
\Upsilon_R^{(k)}(\omega)=<J_k(-\omega)J_k(\omega)>=\frac{B_k(\omega)}{A_k(\omega)}.
\end{equation}
Setting $A_k(\omega)=0$, one can find the location of the poles in the Green function. In such a case, we have $\omega=\operatorname{Re}(\omega)+\operatorname{Im}(\omega)$ in which it has real part (normal mode) and imaginary part (quasi-normal mode).
The imaginary part, which is negative, introduces inverse relaxation time $\tau_d$ of the desired mode as $\omega = \operatorname{Re}[\omega] + i\operatorname{Im}[\omega] \equiv \operatorname{Re}[\omega] -i2\pi / \tau_d$. Moreover, it describes exponential damping with a characteristic time scale set by $\tau_d$.
\\
On the other hand, the quasi-normal modes play the role of controller for the black hole ringdown in order to decay or collapse perturbed black holes towards their hairless cases or a relaxation towards thermal equilibrium after a perturbation period. Also, the quasi-normal modes with the smallest value of $\operatorname{Im}[\omega]$ have a less damped state and supervise to overlook the thermalization time- scale. Since the black hole ringdown is dual to the thermalization process of the CFT \cite{Horowitz:1999jd}, the quasi-normal modes appear as the poles of the Fourier transform of the retarded Green's function in the context of the Ruelle resonances \cite{Birmingham:2001pj}.
\\
In the context of the WGC, we encounter $\tau_d>0$. Because the negative values of $\tau_d$ correspond to an exponentially-growing unstable fundamental state, which threatens the stability of the geometry of black holes. Besides the positive value, it should have a specific bound because the thermalization process can not be happened arbitrarily fast. Hence, we engage a conjecture \cite{Urbano:2018kax}  
\begin{equation}\label{eq3}
\tau_d\geq \frac{1}{T},  
\end{equation}
see Refs.\cite{Horowitz:1999jd,Hod:2006jw,Hod:2017uqc,Sachdev:1992fk,Maldacena:2016hyu,Eberlein:2017wah,Gruzinov:2007ai} for more discussion. Due to the perturbations with different thermalization time scales, the key point is that the perturbations decay very fast in order to satisfy the condition $\tau_d >c/T$. From exiting the perturbations from the metric background, this could be realized that for a near-extremal charged black hole, there must be at least one particle to fulfill the thermalization rate $\tau_d > c/T$.
\\
In fact, considering a particle near a black hole leads to producing some perturbations that could be described by the Klein-Gordon equation (\ref{eq1}) as the Equation of Motion of the particle. By solving this equation, we find an imaginary value of energy for the particle which is connected to eq.(\ref{eq3}). 
On the other hand, the existence of such perturbations makes the black holes unstable so that they approach the extremal limit and consequently connect to the WGC condition. By using the mentioned relationship between eq.(\ref{eq3}) and the WGC and also the polarity quantity $w$ that we obtain during our calculations, we can study the WGC condition from the universal equation (\ref{eq3}) (see Refs.\cite{Urbano:2018kax,Hod:2017uqc,Hod:2010hw}).

In the next section, we use the above description for two black holes, e.g., HSV and KNA.

\section{Hyperscaling violating black holes}\label{hsv}

The metric of the HSV black hole with mass $M$ and charge $Q$ is given by \cite{Urbano:2018kax,Pedraza:2018eey,Hod:2017uqc,Horowitz:1999jd}
\begin{equation}\label{eq4}
ds^2=r^{\frac{-2\theta}{\textrm{d}}}\left[-r^{2\textrm{z}}f(r)dt^2+\frac{ dr^2}{r^2f(r)}+ r^2 d\Omega^2_{k,\textrm{d}}\right].
\end{equation}

Where $z$ and $\theta$ are dynamical and Hyperscaling violation parameters, respectively. Also, $f(r)$  takes the form
\begin{equation}\label{eq5}
f(r)=1+\frac{k}{r^2}\frac{(\textrm{d}-1)^2}{(\textrm{z}+\textrm{d}-\theta-2)^2}-\frac{M}{r^{\textrm{z}+\textrm{d}-\theta}}+
\frac{Q^2}{r^{2(\textrm{z}+\textrm{d}-\theta)}}.
\end{equation}

Here, $k=-1, 0, 1$ determines the hyperboloid, planar and spherical topology for the black hole horizon \cite{Pedraza:2018eey}. In this work, we consider the $k=0$, so we have
\begin{eqnarray}\label{eq6}
ds^2=r^{\frac{-2\theta}{\textrm{d}}}\left[-r^{2\textrm{z}}f(r)dt^2+\frac{ dr^2}{r^2f(r)}+ r^2 d\vec{x}^2\right],\hspace{1cm}f(r)=1-\frac{M}{r^{\textrm{z}+\textrm{d}-\theta}}+\frac{Q^2}{r^{2(\textrm{z}+\textrm{d}-\theta)}},
\end{eqnarray}
where $d\vec{x}^2=\sum_{i=1}^{d}dx_i^2$ and $x_i$ are spatial coordinates of a $d$ dimensional space. By setting $f(r_H) = 0$, we obtain the event horizon radius, $r_H$, for a charged black hole solution
\begin{equation}\label{eq7}
r_H^{2(\textrm{d}+\textrm{z}-\theta-1)}-M r_H^{\textrm{d}+\textrm{z}-\theta-2}+Q^2=0.
\end{equation}
Using the formula $T=\frac{r_H^{z+1}}{4\pi}|\acute{f}(r_H)|$ and  \eqref{eq6}, one can obtain the Hawking temperature \cite{Alishahiha:2012qu,Caldarelli:1999xj}
\begin{equation}\label{eq8}
T=\frac{(\textrm{z}+\textrm{d}-\theta)r_H^\textrm{z}}{4\pi}\left(1-\frac{(\textrm{z}+\textrm{d}-\theta-2)Q^2}{\textrm{z}+\textrm{d}-\theta}
r_H^{2(-\textrm{z}-\textrm{d}+\theta+1)}\right).
\end{equation}
Therefore, by putting $T=0$, the extremality bound is obtained by
\begin{equation}\label{eq9}
r_H^{2(\textrm{z}+\textrm{d}-\theta-1)}=\frac{(\textrm{z}+\textrm{d}-\theta-2)}{\textrm{z}+\textrm{d}-\theta} Q^2.
\end{equation}
In this case, we use relations \eqref{eq7} and \eqref{eq9} and rewrite $f(r)$ in terms of $r_H$
\begin{equation}\label{eq10}
f(r)=1-\frac{2(\textrm{z}+\textrm{d}-\theta-1)}{\textrm{z}+\textrm{d}-\theta-2}\left(\frac{r_H}{r}\right)^{\textrm{z}+\textrm{d}-\theta}+
\frac{\textrm{z}+\textrm{d}-\theta}{\textrm{z}+\textrm{d}-\theta-2}\left(\frac{r_H}{r}\right)^{2(\textrm{z}+\textrm{d}-\theta-1)}.
\end{equation}
Also the potential is given by \cite{Pedraza:2018eey}
\begin{equation}\label{eq11}
A_t=\frac{\sqrt{2(\textrm{z}+\textrm{d}-\theta)(d-\theta)}}{\textrm{z}+\textrm{d}-\theta-2} r_H^{\textrm{z}+\textrm{d}-\theta-1}\left(\frac{1}{r^{\textrm{z}+\textrm{d}-\theta-2}}-\frac{1}
{r_H^{\textrm{z}+\textrm{d}-
\theta-2}}\right).
\end{equation}
By changing the corresponding  coordinates and examining $r$ near the event horizon, we can write the geometry of $AdS_2\times R^{d-1}$ \cite{Alishahiha:2012qu}
 \begin{equation}\label{eq12}r=r_H+\frac{\epsilon r_H^2}{(\textrm{z}+\textrm{d}-\theta)(\textrm{z}+\textrm{d}-\theta-1)\zeta},\hspace{0.5cm} t=\frac{\tau}{\epsilon r_H^\textrm{z}}.
\end{equation}
Here we note that equations \eqref{eq12},\eqref{eq10} and \eqref{eq6} are obtained by the limit $\epsilon \rightarrow 0$ as follows
\begin{equation}\label{eq13}
ds^2=r_H^{2-\frac{2\theta}{\textrm{d}}}\left[\frac{-d\tau^2+d\zeta^2}{(\textrm{z}+\textrm{d}-\theta)(\textrm{z}+\textrm{d}
-\theta-1)\zeta^2}+d\vec{x}^2\right].
\end{equation}
Also, we will have
\begin{equation}\label{eq14}
\begin{split}
A_\tau=\frac{\sqrt{2(\textrm{z}+\textrm{d}-\theta)(\textrm{d}-\theta)}}{(\textrm{z}+\textrm{d}-\theta)(\textrm{z}+\textrm{d}-\theta-1)\zeta}\times r_H^{-\textrm{z}+2}
\end{split}.
\end{equation}
Now, using \eqref{eq1}, \eqref{eq13} and \eqref{eq14} also with respect to that fact that metric allows for the separation of variables, $\Phi(\tau,\zeta,\vec{x})=e^{-i\omega \tau}e^{i\vec{k}.\vec{x}} \phi(\zeta)$, one can calculate
\begin{equation}\label{eq15}
\begin{split}
\partial_{\zeta}^2\phi(\zeta)+\left(\omega+\frac{qr_H^{-z+\textrm{2}}\sqrt{2(\textrm{z}+\textrm{d}-\theta)(\textrm{d}-\theta)} }{(\textrm{z}+\textrm{d}-\theta)(\textrm{z}+\textrm{d}-\theta-1)\zeta}\right)^2\phi(\zeta)-\frac{k^2+m^2 r_H^{2(1-\frac{\theta}{d})}}{(\textrm{z}+\textrm{d}-\theta)(\textrm{z}+\textrm{d}-\theta-1)\zeta^2}\phi(\zeta)=0
\end{split}.
\end{equation}
According to the above equation, $-k^2$ is the eigenvalue of the Laplacian in the flat base sub-manifold. Concerning the above equations, $\phi(\zeta) $ is obtained in terms of the Whittaker functions, Which are given by
\begin{equation}\label{eq16}
\begin{split}
\phi(\zeta)=&c_1 Whittaker M\left[\frac{-iqr_H^{-\textrm{z}+2}\sqrt{2(\textrm{z}+\textrm{d}-\theta)(d-\theta)} }{(\textrm{z}+\textrm{d}-\theta)(\textrm{z}+\textrm{d}-\theta-1)},\nu_k,2ir\omega\right]\\
&+c_2 Whittaker W\left[\frac{-iq r_H^{-\textrm{z}+2}\sqrt{2(\textrm{z}+\textrm{d}-\theta)(\textrm{d}-\theta)}}{(\textrm{z}+\textrm{d}-\theta)(\textrm{z}+\textrm{d}-\theta-1)},\nu_k,2ir\omega\right].
\end{split}
\end{equation}
Also, with respect to the above equation, the $\nu_k$ is defined by
\begin{eqnarray}\label{eq17}
&\nu_k=\sqrt{\frac{1}{4}+\frac{k^2}{(\textrm{z}+\textrm{d}-\theta)(\textrm{z}+\textrm{d}-\theta-1)}-\frac{q^2 r_H^{-\textrm{2z}+4} 2(\textrm{z}+\textrm{d}-\theta)(\textrm{d}-\theta) }{(\textrm{z}+\textrm{d}-\theta)^2(\textrm{z}+\textrm{d}-\theta-1)^2}+m^2 r_H^{2(1-\frac{\theta}{\textrm{d}})}}.
\end{eqnarray}
We want to examine the obtained function  $\phi(\zeta)$  near the event horizon. To do this, we use the limit $\zeta \longrightarrow 0$. Given the properties of the Whittaker function and the definition of $\Delta_k=\frac{1}{2}-\nu_k$, we will have
\begin{equation}\label{eq18}
\begin{split}
\phi(\zeta)_{\zeta \rightarrow 0}\approx & C \bigg[\frac{\Gamma(2\nu_k)(-2i\omega)^{\frac{1}{2}-\nu_k}}{\Gamma\bigg(\frac{1}{2}+\nu_k-\frac{iq r_H^{-\textrm{z}+2} \sqrt{2(\textrm{z}+\textrm{d}-\theta)(\textrm{d}-\theta)}}{(\textrm{z}+\textrm{d}-\theta)(\textrm{z}+\textrm{d}-\theta-1)}\bigg)}\zeta^{{\frac{1}{2}-\nu_k}}+\frac{\Gamma(-2\nu_k)(-2i\omega)^{\frac{1}{2}+\nu_k}}{\Gamma\bigg(\frac{1}{2}-\nu_k-\frac{iq r_H^{-\textrm{z}+2}\sqrt{2(\textrm{z}+\textrm{d}-\theta)(\textrm{d}-\theta)} }{(\textrm{z}+\textrm{d}-\theta)(\textrm{z}+\textrm{d}-\theta-1)}\bigg)}\zeta^{{\frac{1}{2}+\nu_k}}\bigg]\\
&\equiv B_k(\omega)\zeta^{\Delta_k}+A_k(\omega)\zeta^{1-\Delta_k}.
\end{split}
\end{equation}
Using equations \eqref{eq2} and \eqref{eq18}, the correlation function is obtained by following equation
\begin{equation}\label{eq19}
\begin{split}
&\mathcal{X}=\Gamma\big(2\nu_k\big)\Gamma\bigg(\frac{1}{2}-\nu_k-\frac{iq r_H^{-\textrm{z}+2} \sqrt{2(\textrm{z}+\textrm{d}-\theta)(\textrm{d}-\theta)} }{(\textrm{z}+\textrm{d}-\theta)(\textrm{z}+\textrm{d}-\theta-1)}\bigg)\\
&\mathcal{Y}=\Gamma\big(-2\nu_k\big)\Gamma\bigg(\frac{1}{2}+\nu_k-\frac{iq r_H^{-\textrm{z}+2}\sqrt{2(\textrm{z}+\textrm{d}-\theta)(\textrm{d}-\theta)} }{(\textrm{z}+\textrm{d}-\theta)(\textrm{z}+\textrm{d}-\theta-1)}\bigg)\\
&\Upsilon_R^{(k)}(\omega)=(2\omega)^{-2\nu_k}e^{i\pi\nu_k}\times\frac{\mathcal{X}}{\mathcal{Y}}
\end{split}.
\end{equation}
According to equation \eqref{eq19}, do not have $\omega$ with an imaginary part; as a result, WGC can not be discussed.
Now we consider condition $r_H \rightarrow r_H+\frac{\epsilon r_H^2}{(\textrm{z}+\textrm{d}-\theta)(\textrm{z}+\textrm{d}-\theta-1)\zeta_0}$ with respect to \eqref{eq13} and \eqref{eq14}, so we can obtain
\begin{equation}\label{eq20}
ds^2=\frac{r_H^{2-\frac{2\theta}{\textrm{d}}}}{(\textrm{z}+\textrm{d}-\theta)(\textrm{z}+\textrm{d}-\theta-1)\zeta^2}\left[-(1-\frac{\zeta^2}{\zeta_0^2})d\tau^2+
(1-\frac{\zeta^2}{\zeta_0^2})^{-1}d\zeta^2\right]
+r_H^{2-\frac{2\theta}{\textrm{d}}}d\vec{x}^2,
\end{equation}
and
\begin{equation}\label{eq21}
A_\tau=\frac{\sqrt{2(\textrm{z}+\textrm{d}-\theta)(\textrm{d}-\theta)}}{(\textrm{z}+\textrm{d}-\theta)(\textrm{z}+\textrm{d}-\theta-1)\zeta} r_H^{-\textrm{z}+2}\left(1-\frac{\zeta}{\zeta_0}\right).
\end{equation}
The temperature associated to the metric in equation \eqref{eq20} is $T = 1/2\pi \zeta_0$. Using \eqref{eq1}, \eqref{eq20} and \eqref{eq21} also with respect to this issue that the metric allows for the separation of variables, $\Phi(\tau,\zeta,\vec{x})=e^{-i\omega \tau}e^{i\vec{k}.\vec{x}} \phi(\zeta)$, one can calculate
\begin{equation}\label{eq22}
\begin{split}
&\partial_{\zeta}^2\phi(\zeta)+\frac{2\zeta}{\zeta^2-\zeta_0^2}\partial_{\zeta}\phi(\zeta)+
\frac{\left[\omega+\frac{q r_H^{-\textrm{z}+2} \sqrt{2(\textrm{z}+\textrm{d}-\theta)(\textrm{d}-\theta)} (1-\frac{\zeta}{\zeta_0})}{(\textrm{z}+\textrm{d}-\theta)(\textrm{z}+\textrm{d}-\theta-1)\zeta}\right]^2}{(1-\frac{\zeta^2}{\zeta_0^2})^2}\phi(\zeta)\\
&-\frac{k^2+m^2 r_H^{2(1-\frac{\theta}{\textrm{d}})}}{(\textrm{z}+\textrm{d}-\theta)(\textrm{z}+\textrm{d}-\theta-1)\zeta^2(1-\frac{\zeta^2}{\zeta_0^2})}\phi(\zeta)=0.
\end{split}
\end{equation}
By solving the above equation, $\phi(\zeta,\omega)$ is obtained by
\begin{equation}\label{eq23}
\begin{split}
&\phi_k(\zeta,\omega)\sim
\left(\frac{1}{\zeta}-\frac{1}{\zeta_0}\right)^{-\frac{1}{2}\mp\nu_k}\left(\frac{\zeta_0+\zeta}{\zeta_0-\zeta}\right)^{i\omega \zeta_0-i\frac{q  r_H^{-\textrm{z}+2} \sqrt{2(\textrm{z}+\textrm{d}-\theta)(\textrm{d}-\theta)}}{(\textrm{z+d}-\theta)(\textrm{z+d}-\theta-1)}}\\
&\times F_1\bigg\{\frac{1}{2}\pm \nu_k+i\omega \zeta_0-i\frac{q \sqrt{2(\textrm{z}+\textrm{d}-\theta)(\textrm{d}-\theta)} r_H^{-\textrm{z}+2}}{(\textrm{z}+\textrm{d}-\theta)(\textrm{z}+\textrm{d}-\theta-1)},
\frac{1}{2}\\
&\pm \nu_k+-i\frac{q \sqrt{2(\textrm{z}+\textrm{d}-\theta)(\textrm{d}-\theta)} r_H^{-\textrm{z}+2}}{(\textrm{z}+\textrm{d}-\theta)(\textrm{z}+\textrm{d}-\theta-1)},\frac{1}{2}\pm \nu_k,\frac{2\zeta}{\zeta-\zeta_0}\bigg\}.
\end{split}
\end{equation}
We examine the above solution on the AdS boundary. In that case, using the properties of the hypergeometric functions and the equation \eqref{eq2}, we obtain the retarded Green’s function as follows
\begin{equation}\label{eq24}
\begin{split}
&\mathcal{A}=\Gamma\big(1+2\nu_k\big)\Gamma\bigg(\frac{1}{2}-\nu_k-\frac{i\omega}{2\pi T}+\frac{iq r_H^{-\textrm{z}+2}\sqrt{2(\textrm{z}+\textrm{d}-\theta)(\textrm{d}-\theta)}}{(\textrm{z}+\textrm{d}-\theta)(\textrm{z}+\textrm{d}-\theta-1)}\bigg)\\
&\times\Gamma\bigg(\frac{1}{2}-\nu_k+\frac{iqr_H^{-\textrm{z}+2}\sqrt{2(\textrm{z}+\textrm{d}-\theta)(\textrm{d}-\theta)}}{(\textrm{z}+\textrm{d}-\theta)
(\textrm{z}+\textrm{d}-\theta-1)}\bigg)\\,
&\mathcal{B}=\Gamma\big(1-2\nu_k\big)\Gamma\bigg(\frac{1}{2}+\nu_k-i\frac{\omega}{2\pi T}+\frac{iq r_H^{-\textrm{z}+2}\sqrt{2(\textrm{z}+\textrm{d}-\theta)(\textrm{d}-\theta)}}{(\textrm{z}+\textrm{d}-\theta)(\textrm{z}+\textrm{d}-\theta-1)}\bigg)\\
&\times\Gamma\bigg(\frac{1}{2}+\nu_k+\frac{iqr_H^{-\textrm{z}+2}\sqrt{2(\textrm{z}+\textrm{d}-\theta)(\textrm{d}-\theta)}}{(\textrm{z}+\textrm{d}-\theta)
(\textrm{z}+\textrm{d}-\theta-1)}\bigg),\\
&\Upsilon_R^{(k)}(\omega)=(4\pi T)^{-2\nu_k}\times\frac{\mathcal{A}}{\mathcal{B}}
\end{split}.
\end{equation}
We used the explicit definition of $T$ instead of $\zeta_0$. When we get the polarity of Equation \eqref{eq2}, $\omega$ has two parts, real and imaginary as $\omega=Re(\omega)+Im(\omega)$.
\begin{equation}\label{eq25}
\omega=\frac{q r_H^{-\textrm{z}+2}\sqrt{2(\textrm{z}+\textrm{d}-\theta)(\textrm{d}-\theta)}}{(\textrm{z}+\textrm{d}-\theta)(\textrm{z}+\textrm{d}-\theta-1)}-i2\pi T\left(\frac{1}{2}+n+\nu_k\right).
\end{equation}
Here, we can talk about WGC because we have a quasi-normal mode. So we note here that we are interested in the system's response at low frequencies \cite{Hod:2010hw,Hod:2006jw,Horowitz:1999jd}.
\begin{equation}\label{eq26}
\tau_d \equiv \tau_d^{(0,0)}= \frac{1}{T(\frac{1}{2}+\nu_0)},\hspace{0.5cm}  \nu_0=\sqrt{\frac{1}{4}-\frac{2q^2 r_H^{-2\textrm{z}+4}(\textrm{z}+\textrm{d}-\theta)(\textrm{d}-\theta)}{(\textrm{z}+\textrm{d}-\theta)^2(\textrm{z}+\textrm{d}-\theta-1)^2}+m^2 r_H^{2(1-\frac{\theta}{\textrm{d}})}}.
\end{equation}
Now using equations \eqref{eq3} and \eqref{eq26}, we get the following condition
\begin{equation}\label{eq27}
\left(m r_H^{1-\frac{\theta}{\textrm{d}}}-\frac{q r_H^{-\textrm{z}+2} \sqrt{2(\textrm{z}+\textrm{d}-\theta)(\textrm{d}-\theta)}}{(\textrm{z}+\textrm{d}-\theta)(\textrm{z}+\textrm{d}-\theta-1)}\right)\left(m r_H^{1-\frac{\theta}{\textrm{d}}}+\frac{q r_H^{-\textrm{z}+2} \sqrt{2(\textrm{z}+\textrm{d}-\theta)(\textrm{d}-\theta)}}{(\textrm{z}+\textrm{d}-\theta)(\textrm{z}+\textrm{d}-\theta-1)}\right)<0.
\end{equation}
The WGC condition is satisfied when we have the following expression
\begin{equation}\label{eq28}
r_H^{1-\frac{\theta}{\textrm{d}}}=\frac{ r_H^{-\textrm{z}+2} \sqrt{2(\textrm{z}+\textrm{d}-\theta)(\textrm{d}-\theta)}}{(\textrm{z}+\textrm{d}-\theta)(\textrm{z}+\textrm{d}-\theta-1)}.
\end{equation}
According to eq.(\ref{eq27}), we can rewrite the radius of the event horizon in terms of dynamic parameters as follows,
\begin{equation}\label{new29}
r_H^{z-\frac{\theta}{d}-1}=\frac{  \sqrt{2(z+d-\theta)(d-\theta)}}{(z+d-\theta)(z+d-\theta-1)}.
\end{equation}
This relation shows the compatibility of the WGC and the CFT only on the horizon of a specific event obtained from eq.(\ref{new29})
According to the relation \eqref{eq9}, $r$ is positive when we have relation as $\textrm{z}+\textrm{d}-\theta>2$. The answer to the above relationship is when we have $\textrm{z}=1 , \textrm{d}=1$ and $\theta\rightarrow 0^-$ or $\textrm{z} \rightarrow1^+ , \textrm{d}=1$  and $ \theta =0$.\\

To drive the WGC from the CFT viewpoint, we need to specify a series of specific points and discuss the compatibility of these two structures in those points. Since $r_H$ represents the position of the whole horizon, we can not discuss the relation between the WGC of the CFT for the entire horizon parameters. Therefore, we must inevitably determine a specific value for $r_{H}$ by setting some important parameters such as z, d, and $\theta$ in order to find specific points of $r_H$. Consequently, we can say that by setting $r_H<1$ and for a series of specific points, we can say the WGC and the CFT are aligned with each other.
\begin{figure}[tbh]
\hspace*{1cm}
\begin{center}
\includegraphics[width=.45\textwidth,keepaspectratio]{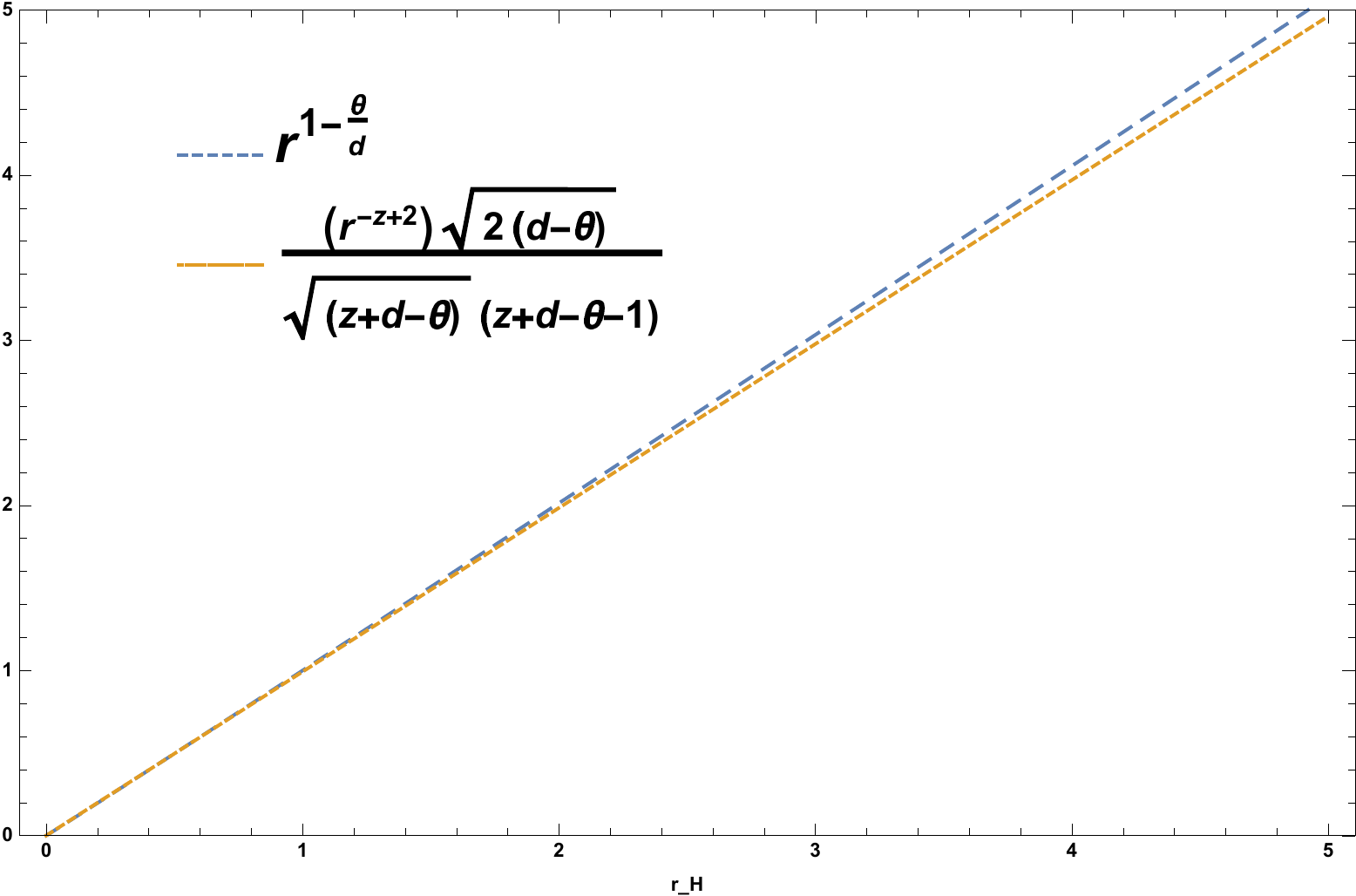}
\includegraphics[width=.45\textwidth,keepaspectratio]{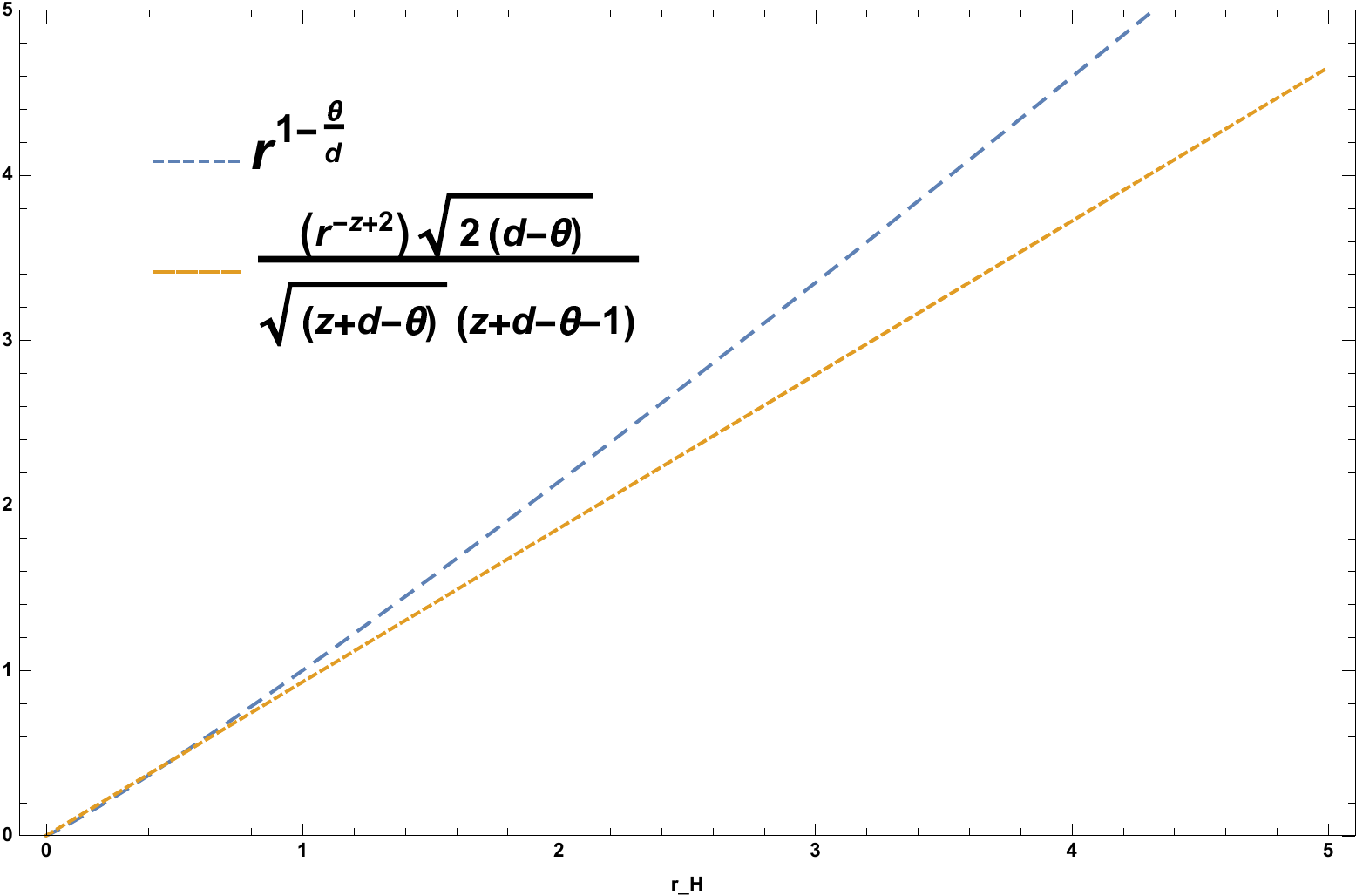}
\caption{\small{Left plot:  $z=1$ , $ d = 1$, and  $\theta = -0.01$
 Right plot:  $z=1$ , $ d = 1$, and  $\theta = -0.1$ }}
 \label{fig1}
\end{center}
\end{figure}
As we can see in fig. \ref{fig1}, when $\theta$ is closer to zero, WGC holds for bigger values of $r_H$. In that case, it is set for both $r_H$ greater and less than one.
\begin{figure}[tbh]
\hspace*{1cm}
\begin{center}
\includegraphics[width=.45\textwidth,keepaspectratio]{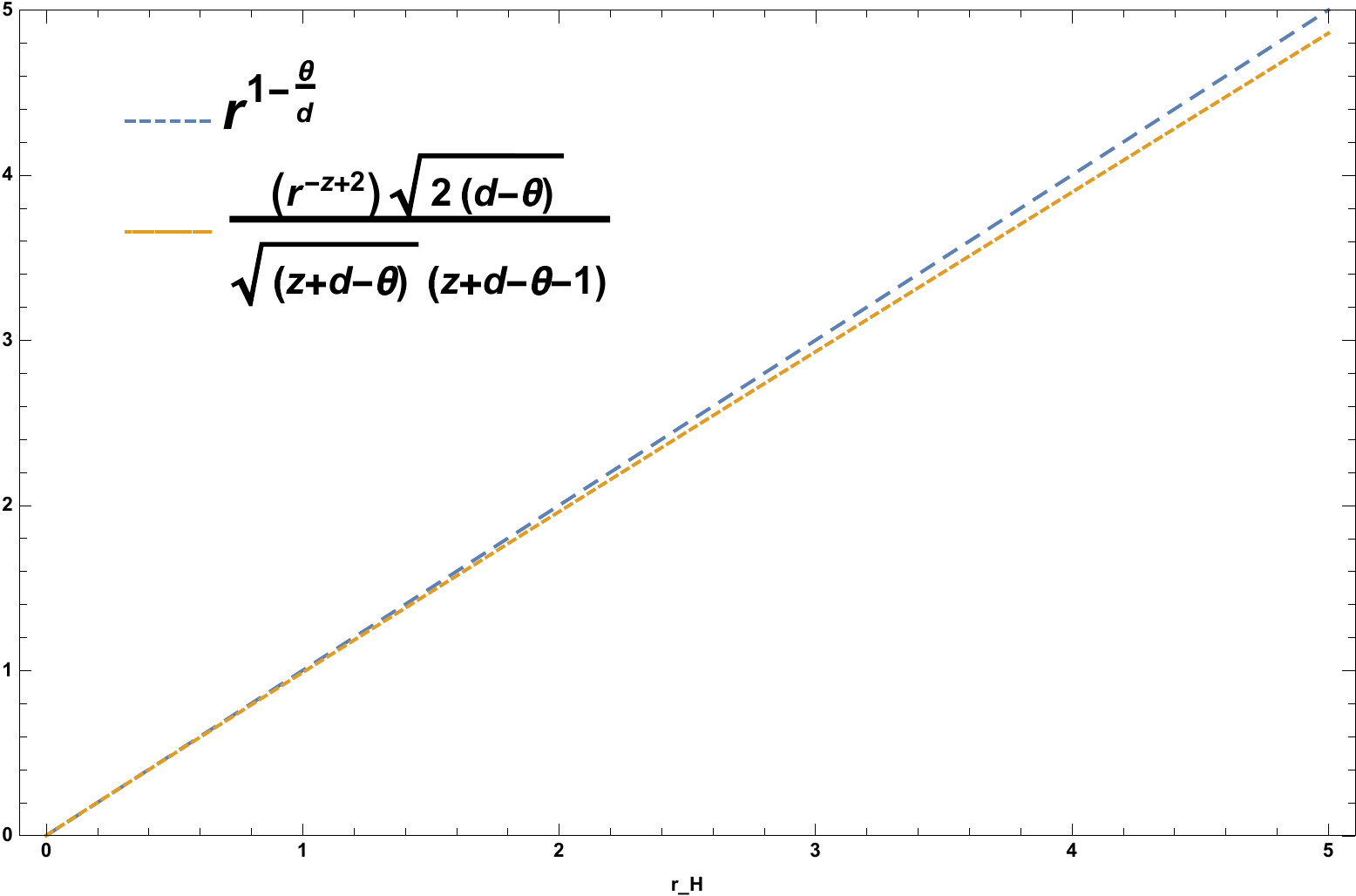}
\includegraphics[width=.45\textwidth,keepaspectratio]{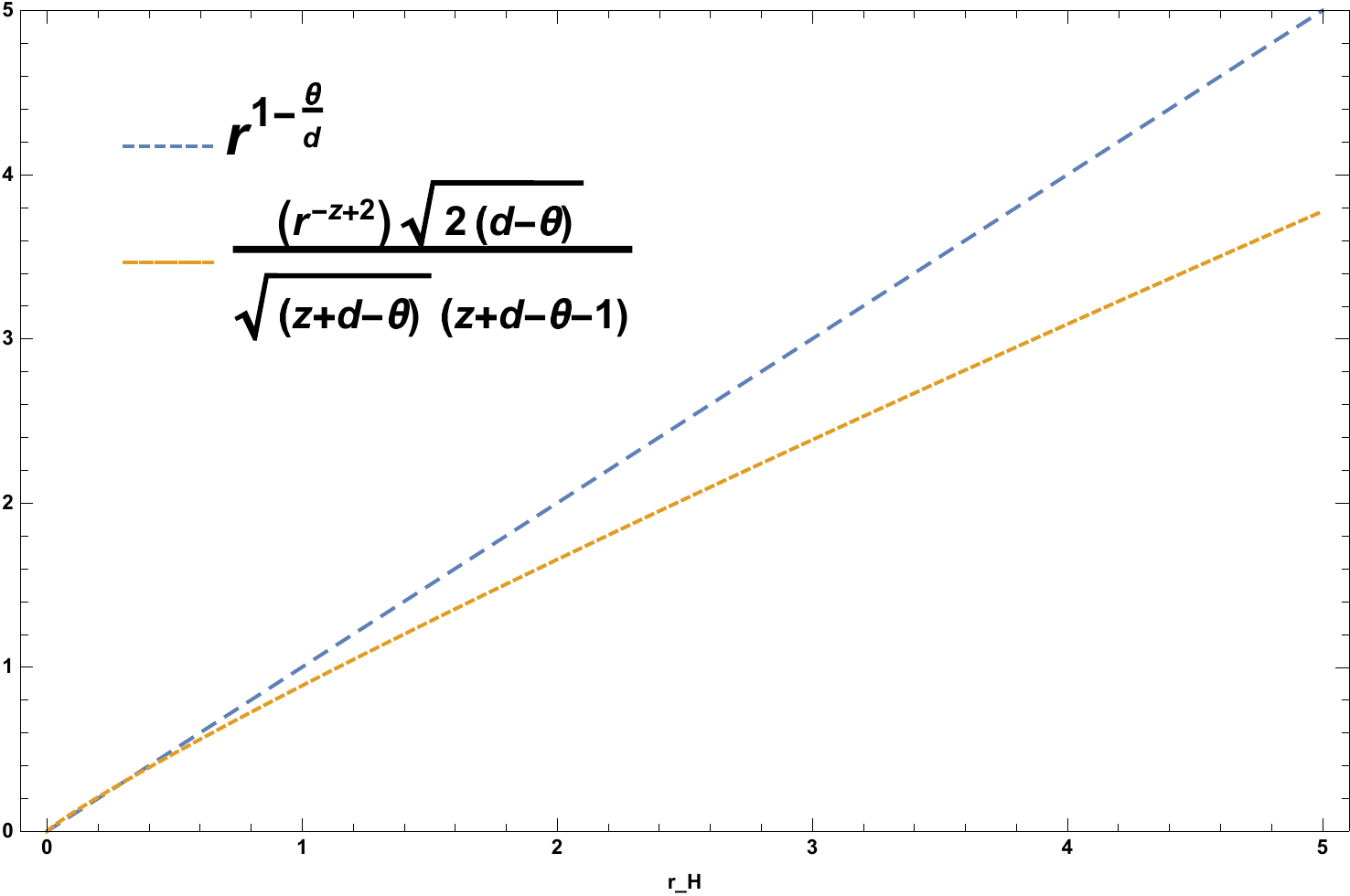}
\caption{\small{Left plot:  $z=1.01$ , $ d = 1$, and  $\theta = 0$
 Right plot:  $z=1.1$ , $ d = 1$, and  $\theta = 0$ }}
 \label{fig2}
\end{center}
\end{figure}
In fig \ref{fig2}, also we see that when for $\textrm{z}$ that is closer to one, $r_H$ can be established for different values, but $\textrm{z}$ is slightly larger than one, WGC for $r_H<1$ will be established. When $\theta<\textrm{d}$ we have an solution for a value of $r_H<1$.
\begin{figure}[tbh]
\hspace*{1cm}
\begin{center}
\includegraphics[width=.45\textwidth,keepaspectratio]{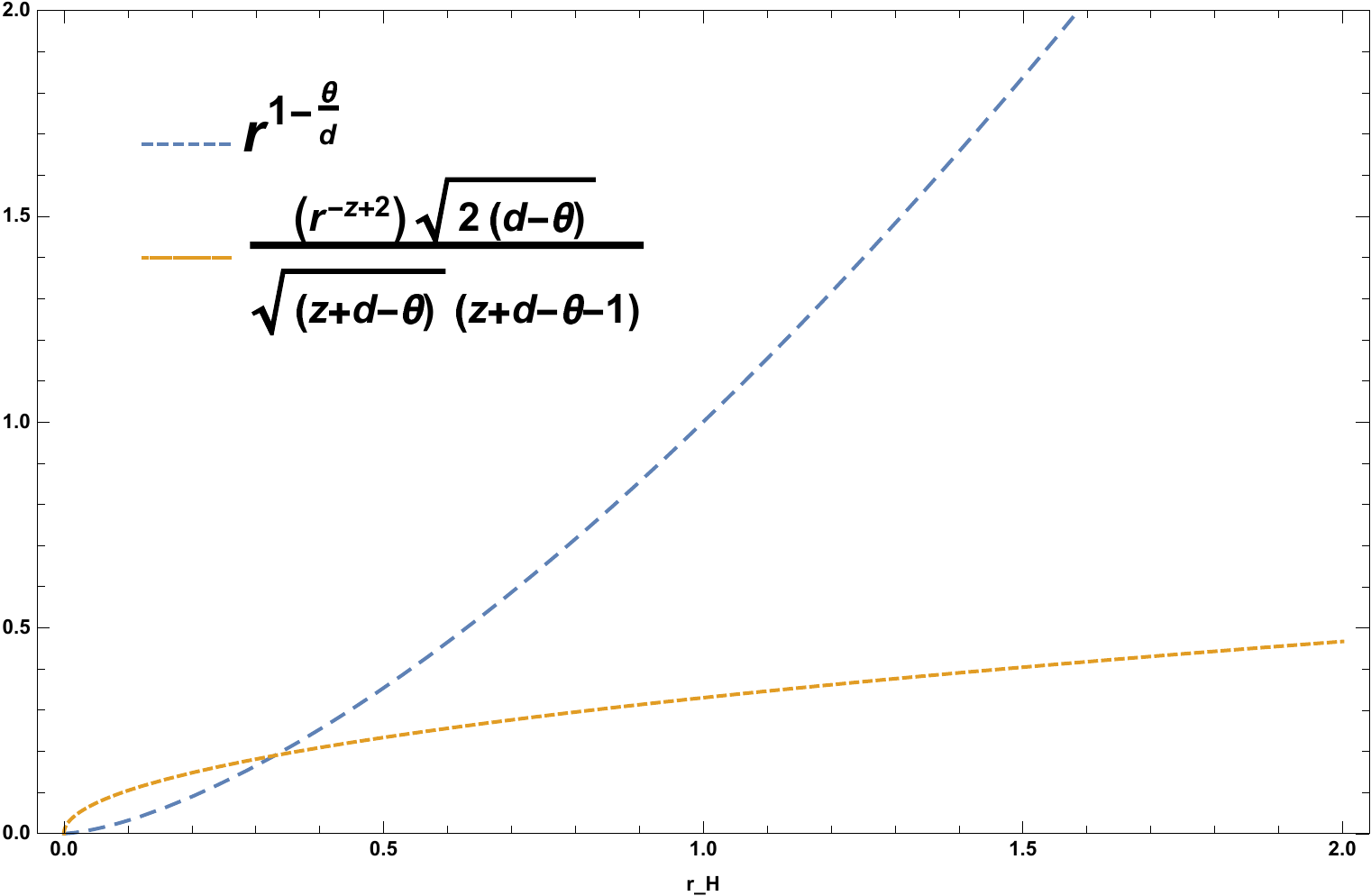}
\includegraphics[width=.45\textwidth,keepaspectratio]{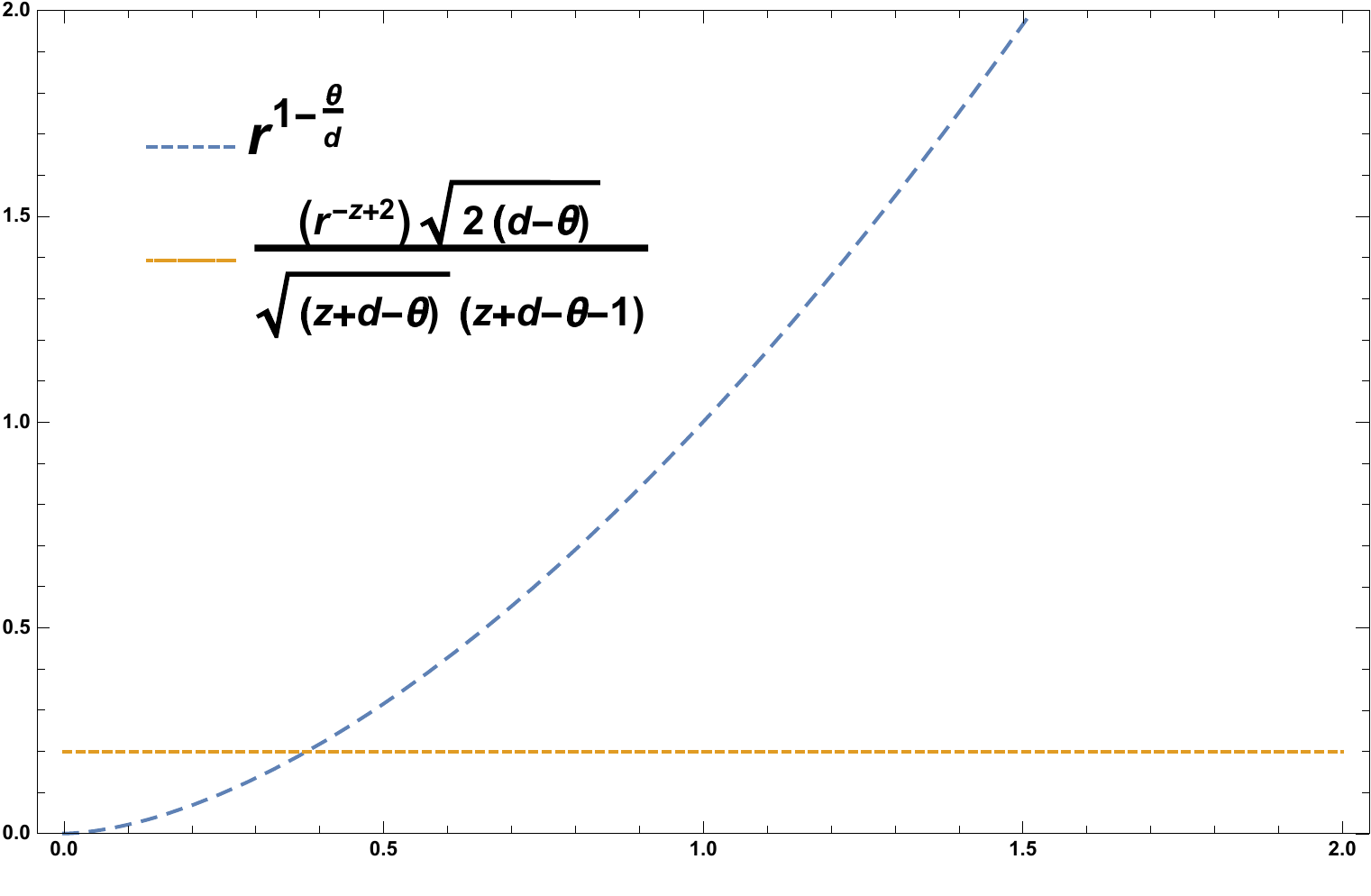}
\caption{\small{Left plot:  $z=1.5$ , $ d = 2$, and  $\theta = -1$
 Right plot:  $z=2$ , $ d = 3$, and  $\theta = -2$ }}
 \label{fig3}
\end{center}
\end{figure}
From fig \ref{fig3}, when we consider $\theta $ to be a number less than zero and $\textrm{z}$ and $\textrm{d}$ to be greater than one, then WGC will hold for a value of $r_H<1$.

\section{Kerr-Newman-AdS black holes}\label{knb}
Analogous to the previous black hole, we attempt to obtain the wave function and the energy spectrum of the KNA black hole. Here, we find that our black hole is unstable, which is proof of WGC. To see such condition for the corresponding black hole, we consider the Boyer-Lindquist-type coordinates  \cite{Caldarelli:1999xj},
\begin{equation}\label{eq29}
ds^2=-\frac{f(r)}{\rho^2}\left(dt-\frac{a \sin^2\theta}{\Xi}d\phi\right)^2+\frac{\rho^2}{f(r)}dr^2+\frac{\rho^2}{f(\theta)}d\theta^2
+\frac{f(\theta)}{\rho^2}\sin^2\theta\left(adt-\frac{(r^2+a^2)}{\Xi}d\phi\right)^2,
\end{equation}
where
\begin{eqnarray}\label{eq30}
f(r)={(r^2+a^2)(1+\frac{r^2}{\ell^2})-2Mr+Q^2} ,\hspace{0.75cm}
f(\theta)={1-\frac{a^2}{\ell^2}\cos^2\theta},\hspace{0.75cm}
\rho^2={r^2+a^2 \cos^2\theta},\hspace{0.75cm}
\Xi={1-\frac{a^2}{\ell^2}}.
\end{eqnarray}
Here, $a$ is the rotational parameter, and $Q$ is the electric charge. $\ell^2$ is related to the cosmological constant. If it is positive, we have $dS$; when it is negative, we have $AdS$. Its vector potential is considered as follows
\begin{equation}\label{eq31}
A=-\frac{Qr}{\rho^2}\left(dt-\frac{a \sin^2 \theta}{\Xi}d\phi \right).
\end{equation}
The Hawking temperature, entropy, and angular velocity of the horizon are obtained by
\begin{eqnarray}\label{eq32}
{T_H}={\frac{r_+ (1+\frac{a^2}{\ell^2}+\frac{3 r_+^2}{\ell^2}-\frac{a^2+Q^2}{r_+^2})}{4 \pi (r_+^2+a^2)}} ,\hspace{0.75cm}
S={\frac{\pi(r_+^2+a^2)}{\Xi}},\hspace{0.75cm}
{\Omega_H}&=&{\frac{a\Xi}{r_+^2+a^2}}.
\end{eqnarray}
We can rewrite $f(r)$ near the event horizon in quadratic order as follows \cite{Chen:2010bh}
\begin{equation}\label{eq33}
f(r)\simeq k(r-r_+)(r-r_*), \hspace{1cm} k=1+\frac{a^2}{\ell^2}+\frac{6r_+^2}{\ell^2},
\end{equation}
Where $r_+$ is the radius of the external event horizon and $r_*$ is the radius of another horizon. In the extremality limit, the following conditions also apply. \cite{Chen:2010bh}
\begin{equation}\label{eq34}
r_+^2\geq\frac{\ell^2}{6}\left(\sqrt{(1+\frac{a^2}{\ell^2})^2+12\frac{a^2}{\ell^2}}-(1+\frac{a^2}{\ell^2})\right).
\end{equation}
In the limit $\frac{a}{\ell}\ll 1$ we can consider the minimum value of $r_+$ equal to $a$. By putting $f(r)=0$ and $T_H=0$ we can get the extreme state of the black hole with the condition $r_+=a$ as follows
\begin{equation}\label{eq35}
k=\frac{M_{exe}}{r_+}+\frac{4r_+^2}{\ell^2}.
\end{equation}
We now use \eqref{eq1} to solve Klein Gordon's equation in the background of the KNA black hole and obtain the correlation function, which is calculated by the following equation \cite{Chen:2010bh}
\begin{equation}\label{eq36}
\Upsilon_R=\frac{\Gamma(1-2h_Q)}{\Gamma(2h_Q-1)}\times\frac{\Gamma(h_Q+i\frac{\omega_L-Q_L \mu_L}{2\pi T_L})\Gamma(h_Q+i\frac{\omega_R-Q_R \mu_R}{2\pi T_R})}{\Gamma(1-h_Q+i\frac{\omega_L-Q_L \mu_L}{2\pi T_L})\Gamma(1-h_Q+i\frac{\omega_R-Q_R \mu_R}{2\pi T_R})},
\end{equation}
where
\begin{eqnarray}\label{eq37}
{\omega_L}&=&{\frac{r_+^2+r_*^2+2a^2}{2 a \Xi}\omega}, \hspace{2cm}  {\omega_R}={\frac{r_+^2+r_*^2+2a^2}{2 a \Xi}\omega-M},\nonumber\\
{Q_L}&=&Q_R={e} \nonumber\\
{\mu_L}&=&{\frac{Q(r_+^2+r_*^2+2a^2)}{2 a \Xi (r_+ + r_*)}}, \hspace{2cm}    {\mu_R}={\frac{Q(r_+ + r_*)}{2 a \Xi}},\nonumber\\
{T_L}&=&{\frac{k(r_+^2+r_*^2+2a^2)}{4\pi a \Xi (r_+ + r_*)}}, \hspace{2cm}    {T_R}={\frac{k(r_+ - r_*)}{4\pi a \Xi}},
\end{eqnarray}
also we have
\begin{equation}\label{eq38}
h_Q=\frac{1}{2}+\frac{1}{2}\sqrt{1-4\left(\frac{e^2 Q^2}{k^2}-\frac{K_Q}{k}\right)},
\end{equation}
where $K_Q$ is the separation constant \cite{Chen:2010bh}. The pole of equation \eqref{eq36} is given as follows
\begin{equation}\label{eq39}
1-h_Q+i\frac{\omega_L-Q_L \mu_L}{2\pi T_L}=-n,      \hspace{1.5cm}        1-h_Q+i\frac{\omega_R-Q_R \mu_R}{2\pi T_R}=-n.
\end{equation}
Since the above two relations are similar to each other, we consider one of them and place it in relation \eqref{eq3} with the condition of $K_Q=-3k$ and obtain the following equation
\begin{equation}\label{eq40}
\frac{e Q}{k}>1.
\end{equation}
Using equations \eqref{eq34}, \eqref{eq35} and  \eqref{eq40} also with condition $\frac{a}{\ell}\ll 1$, we achieve the following equation
\begin{equation}\label{eq41}
\frac{e a Q}{M_{exe}+\frac{4a^3}{\ell^2}}>1.
\end{equation}
The above statement satisfies WGC condition for $\frac{a}{\ell}\ll 1$ and $e=\frac{1}{a}$.

The initial arguments about the WGC were presented about the physics of black holes and then were followed by some expansions in different parts. 
\\
The primary characteristics of black holes are addressed through the laws of thermodynamics and its relation with gravity based on the AdS/CFT correspondence. In this work, we show the WGC is related to the thermalization dynamics governing the relaxation process after a perturbation process. The validity of such result is guaranteed by having a lower bound on the thermalization time scale. In fact, the CFT gives a lower thermalization bound on the thermalization time-scale to charged-near-extremal black holes, which can be used as a necessary condition to reach the definition of the WGC. By using the model's free parameters, we determine a series of parametric points and regions and then check the WGC in such points. Note that imposing more restrictions might determine the exact range of compatibility between the WGC and the CFT. By considering the common points coming from gravity, thermodynamics and the relationship between gravity and the WGC, the thermodynamics of black holes can be the desired issue in order to enter the WGC and challenge different theories for a better understanding between gauge theories and quantum gravity.

\section {Conclusions and outlook}\label{con}
The idea of WGC has been applied in many cosmological structures, such as inflation, dark energy, and black holes physics. On the other hand,  CFT is engaged a lot in the literature of theoretical particle physics, particularly in the AdS/CFT correspondence. But the connection between the two theories navigates us to some interesting results. The main aim of this paper is to study the relationship between WGC and CFT using HSV and KNA black holes. To fulfill this, we used the correlation function of CFT. Hence, we obtained proof for predicting WGC on the CFT side. This means the emergence of WGC from correlation function calculations in the considered black holes. Since WGC at critical points has compatibility results, it also appears in extremality bound of black holes. Assuming this and the CFT correlation function, we calculated the black holes' critical points. Hence, WGC appeared for both black holes. In addition to proving WGC, we have also shown the exciting close relation between these two ideas, i.e., WGC and CFT. In that case, we used the correlation function in CFT and its poles. Also, we obtained the energy spectrum of the black holes, including two parts, i.e., real (normal mode) and imaginary (quasi-normal mode). We found that when $z=1$, $d=1$ and $\theta\rightarrow 0^{-}$,  WGC emerges in the HSV black holes since it contains $r_{H}$ larger and smaller than one. In other cases, WGC is only valid for $r_H$ less than one. The condition of WGC for the KNA black hole is related to the rotation and radius parameters if the charged particle near the black hole is $\frac{1}{a}$ and has a ratio such as $\frac{a}{\ell}\ll 1$.
Since a relationship between the two ideas has somehow emerged, we can do more calculations to look at the results and even reach a correspondence between the two theories, which we shall examine in future work.
\bibliographystyle{ieeetr}
\bibliography{biblo}
\end{document}